%% file: main.tex
\documentclass[sigconf]{acmart}
\usepackage{hyperref}
\usepackage{wrapfig}
\usepackage{fancybox}
\usepackage{multirow}
\usepackage{rotating}
\usepackage{makecell}
\usepackage{xcolor}
\usepackage{enumitem}
\usepackage[T1]{fontenc}
\usepackage[utf8]{inputenc}
\usepackage{graphicx}
\usepackage{amsmath}
\usepackage{xspace}
\usepackage{balance}

\newcommand{\mymethod}{Mozualization\xspace}

\usepackage{amssymb}

\definecolor{EditedColor}{rgb}{0.2, 0.4, 0.8}

\AtBeginDocument{%
  }

\copyrightyear{2025}
\acmYear{2025}
\setcopyright{rightsretained}
\acmConference[CHI EA '25]{Extended Abstracts of the CHI Conference on Human Factors in Computing Systems}{April 26-May 1, 2025}{Yokohama, Japan}
\acmBooktitle{Extended Abstracts of the CHI Conference on Human Factors in Computing Systems (CHI EA '25), April 26-May 1, 2025, Yokohama, Japan}\acmDOI{10.1145/3706599.3719686}
\acmISBN{979-8-4007-1395-8/2025/04}

\begin{document}

\title{\mymethod: Crafting Music and Visual Representation with Multimodal AI}

\author{Wanfang Xu}
\affiliation{%
  \institution{Xi'an Jiaotong-Liverpool University}
  \city{Suzhou}
  % \state{Jiangsu}
  \country{China}
}
\orcid{0009-0005-3014-2901}

\author{Lixiang Zhao}
\affiliation{%
  \institution{Xi'an Jiaotong-Liverpool University}
  \city{Suzhou}
  % \state{Jiangsu}
  \country{China}
}
\orcid{0000-0001-6181-1673}

\author{Haiwen Song}
\affiliation{%
  \institution{Beijing Institute of Technology}
  \city{Beijing}
  \country{China}
}
\orcid{0009-0000-4386-7024}

\author{Xinheng Song}
\affiliation{%
  \institution{Beijing Institute of Technology}
  \city{Beijing}
  \country{China}
}
\orcid{0000-0002-7930-4079}

\author{Zhaolin Lu}
\affiliation{%
  \institution{Beijing Institute of Technology}
  \city{Beijing}
  % \state{Jiangsu}
  \country{China}
}
\orcid{0009-0008-7775-5352}

\author{Yu Liu}
\affiliation{%
  \institution{Xi'an Jiaotong-Liverpool University}
  \city{Suzhou}
  % \state{Jiangsu}
  \country{China}
}
\orcid{0000-0003-0226-1311}

\author{Min Chen}
\affiliation{%
  \institution{Xi'an Jiaotong-Liverpool University}
  \city{Suzhou}
  % \state{Jiangsu}
  \country{China}
}
\orcid{0000-0002-3122-6788}

\author{Eng Gee Lim}
\affiliation{%
  \institution{Xi'an Jiaotong-Liverpool University}
  \city{Suzhou}
  % \state{Jiangsu}
  \country{China}
}
\orcid{0000-0003-0199-7386}

\author{Lingyun Yu}
\authornote{Corresponding author}
\affiliation{%
  \institution{Xi'an Jiaotong-Liverpool University}
  \city{Suzhou}
  % \state{Jiangsu}
  \country{China}
}
\orcid{0000-0002-3152-2587}

\begin{abstract}
In this work, we introduce \mymethod, a music generation and editing tool that creates multi-style embedded music by integrating diverse inputs, such as keywords, images, and sound clips (e.g., segments from various pieces of music or even a playful cat’s meow).
Our work is inspired by the ways people express their emotions---writing mood-descriptive poems or articles, creating drawings with warm or cool tones, or listening to sad or uplifting music. Building on this concept, we developed a tool that transforms these emotional expressions into a cohesive and expressive song, allowing users to seamlessly incorporate their unique preferences and inspirations. 
To evaluate the tool and, more importantly, gather insights for its improvement, we conducted a user study involving nine music enthusiasts. The study assessed user experience, engagement, and the impact of interacting with and listening to the generated music.
\end{abstract}




\begin{teaserfigure}
 \centering
 \includegraphics[width=0.9\textwidth]{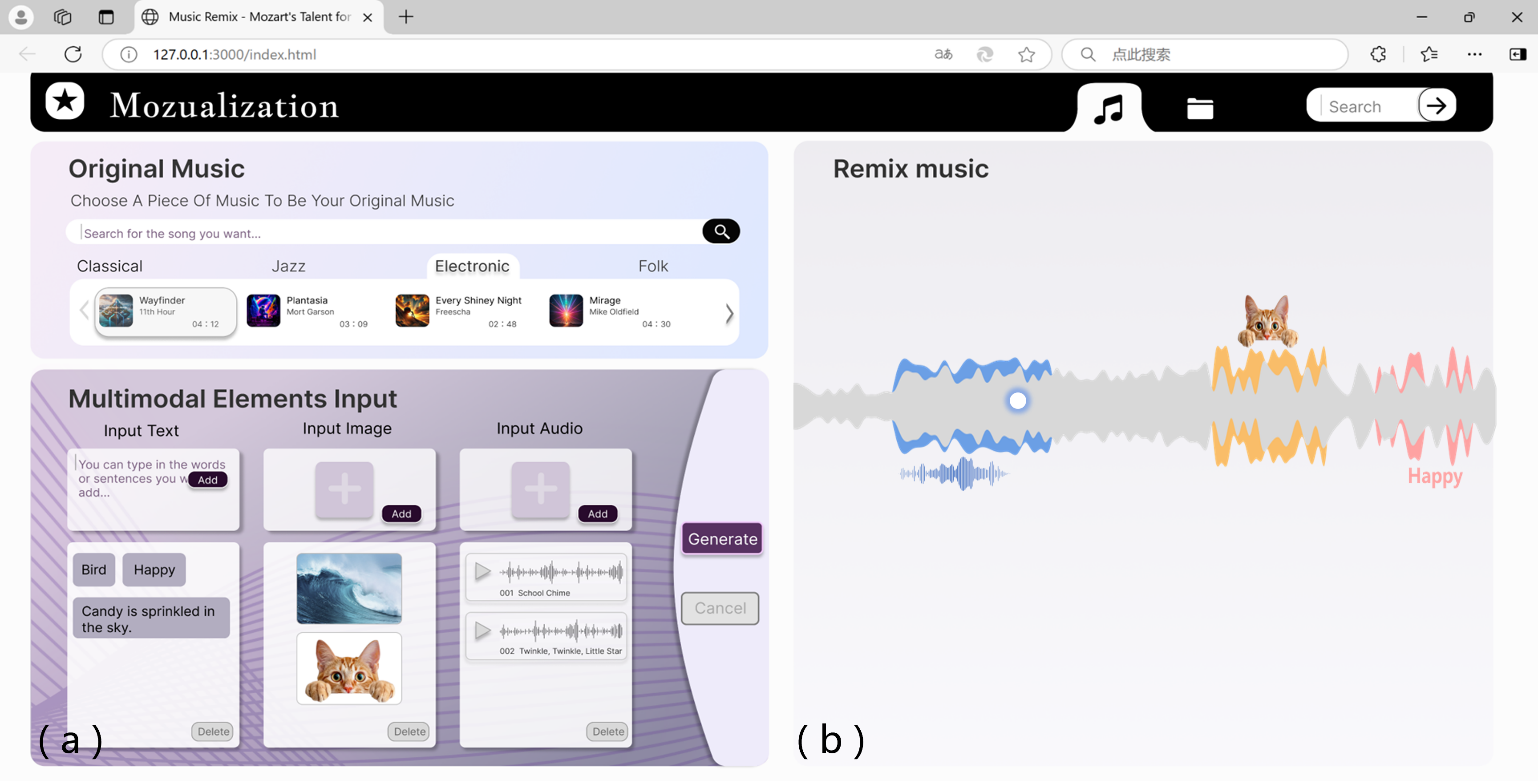}
 \caption{The interface of \mymethod includes a material upload area for text, images, and audio clips (a), and a visualization of the generated music (b).}
 \Description{The interface of \mymethod}
 \label{fig:interface}
 \Description{The interface of Mozualization includes a material upload area for text, images, and audio clips (a), and a visualization of the generated music (b). on the top left is the 'Original Music' selection area, where users can choose music from categories such as classical, jazz, electronic and folk. Below is a multimodal input area where users can enter or upload text, images and audio. On the right side is the visualization area, which shows the original music waveform after mixing and overlays the waveforms of the music elements generated based on the user input, with the original user input text, image and audio near each element waveform.}
\end{teaserfigure}

\maketitle

\input{body/1-Introduction}
\input{body/2-RelatedWork}
\input{body/3-Research}
\input{body/4-SystemDesign}
\input{body/5-UserStudy}
\input{body/6-Result}
\input{body/7-Discussion}
\input{body/8-Conclusion}

\begin{acks}
This work was supported by the National Natural Science Foundation of China (Grant No. 62272396, 52275234), Humanities and Social Sciences Foundation of the Ministry of Education of China (22YJA760055), Science and Technology Innovation Program Project of Beijing Institute of Technology (2024CX01023) and the Fundamental Research Funds for the Central Universities (2024CX06123).
\end{acks}

\bibliographystyle{ACM-Reference-Format}
\balance
\bibliography{reference}

\end{document}

%% file: body/1-Introduction.tex
\section{Introduction}
The use of Artificial Intelligence Generated Content has significantly advanced art and music creation, catering to the growing demand for personalized and customized designs, including music and visual content. This technology benefits users at different skill levels in various ways. By leveraging deep learning and AI algorithms, music creators with prior experience in music composition and editing can enhance their creative process by simulating ideas and exploring different possibilities more efficiently. Meanwhile, music enthusiasts without professional expertise can generate customized music that aligns with their personal designs. Moreover, they can do so without the need to learn complex editing interfaces or acquire in-depth composition knowledge.

The generation framework can be broadly divided into two main stages: \textbf{Music Creation} and \textbf{Music Editing}.
In the creation stage, the main task for AI techniques is to generate foundational musical elements, such as melody, harmony, accompaniment, and rhythm, based on user inputs or predefined parameters. Research has shown that this automation not only accelerates the creative process but also empowers artists to explore new possibilities more efficiently, ultimately inspiring greater creativity~\cite{Wang:2024:TPE,tabak7intelligent,Yang2021ACO}.
In the editing stage, AI technologies are used for refining fine details in existing compositions and augmenting musical elements. Tasks for composers and musicians in this stage include adjusting tempo, modifying instrumentation, adding expressive elements, and integrating additional layers or effects. With AI tools, this process becomes more efficient and interactive, enabling a richer and more personalized editing experience while maintaining the artistic integrity of the piece.

Advanced AI technologies enable music creation and editing not only through traditional tools but also using multimodal inputs such as semantic text and keywords~\cite{Agostinelli:2023:musiclm,lam2024efficient,copet2024simple}. While these methods leverage simple descriptive words to specify the type of music to be generated, they often rely on overly automatic processes, lacking fine-grained control over musical details and melody selection. Consequently, the generated music tends to be standardized, lacking personalization and emotional depth.
Recent research has explored generating music based on images~\cite{bellini2022interactive} and other visual inputs~\cite{kang2024video2music,chowdhury2024melfusion}. These approaches extract themes, colors, and other elements from visual inputs and translate them into musical components. While effective at infusing emotions from diverse forms of input into music, they often fail to clarify how specific visual content influences music composition or how multiple text and visual representations are integrated or merged in the generated results.

To address these challenges, we developed \mymethod, an AI-based music generation system designed to create personalized music using multimodal inputs. Our approach supports diverse ways to \textbf{blend and craft music rich in color, style, and personalization}. \textbf{Color} can be extracted from text that conveys a person's emotions and intentions, and/or from an image filled with warm and cool tones, evoking specific emotional atmospheres. \textbf{Style} can be highly diverse, derived from patterns in images, the structure of melodies, or the genre and type of music. \textbf{Personalization} adds a unique touch by incorporating elements such as a piano melody, a baby's voice, or even the sound of a cat's meow. Our vision is to take these personal and emotional elements, and seamlessly merge them to create a cohesive and expressive piece of music. However, the key challenge lies in how the system can interpret these multimodal inputs and effectively translate them into musical components---such as rhythm, timbre, melody, or harmony---that harmoniously integrate into the composition. Additionally, it is crucial to determine \text{where} and \text{how} these changes should be applied in the original composition to preserve its coherence while enhancing its richness and individuality. The main contributions of our work are:
\begin{itemize}
\item a user-centric design involving professional musicians and novice enthusiasts. 
\item an innovative tool for creating multi-style embedded music by integrating diverse inputs. 
\item a visual representation form to show where and how inputs are applied to enrich the composition. 
\end{itemize}

%% file: body/2-RelatedWork.tex
\section{Related Work}
In this section, we review relevant works on AI-based music creation and editing.

\subsection{AI-based Music Creation}
Music creation focuses on generating music from scratch, including tasks such as notation~\cite{Julia_2018}, recombination sequencing~\cite{Duignan2008ComputerMM}, and symbolic coding for sound synthesis~\cite{Smith01092004}. Integrated music generation tools have been developed to assist composers in structuring their work and creating melodies, harmonies, and rhythms based on predefined musical theories and patterns. Recently, deep learning techniques have been extensively applied to generating music from diverse music corpora~\cite{briot2019deeplearningtechniquesmusic}. As early as 1989, Lewis and Todd ~\cite{todd1989connectionist} proposed the idea of using neural networks for automatic music creation, employing RNNs to sequentially generate music. Since then, various deep learning models have been rapidly developed for music generation. MidiNet ~\cite{yang2017midinet} was designed to learn stylized musical elements from existing melodies for approximate creation, while MuseGAN ~\cite{dong2018musegan} generated polyphonic accompaniments by creating multi-track instrumental sounds from melodies. These approaches have laid a solid technical foundation for music creation. To enhance the creativity and interoperability of music creation, more advanced AI models have been developed to generate music through textual prompts and images. Google's MusicLM~\cite{Agostinelli:2023:musiclm} and Meta's MusicGen ~\cite{copet2024simple} generate music in various styles based on detailed text prompts, while ByteDance's Seed-Music~\cite{bai2024seed} integrates style descriptions, audio references, and music score as multimodal inputs. Additionally, MeLFusion~\cite{chowdhury2024melfusion} synthesizes music by using text descriptions and corresponding images. Although these models provide diverse approaches to music creation, they rely on an automated generation process that lacks transparency. Musicians and composers may find it difficult to grasp how music is generated or how they can participate in and influence the creative process.

\subsection{AI-based Music Editing}
Music editing focuses on modifying and refining existing compositions, including tasks such as audio splicing, pitch and tone adjustment, and applying mixing effects. These tasks typically require musicians to analyze the original material, make creative adjustments, and blend multiple independently recorded tracks, such as instruments and vocals. 
AI techniques have been introduced to assist musicians in these tasks. For example, Suno~\cite{Villapalos2024} employed AI to enable users to splice songs using their own audio input, such as hummed melodies or played chords. 
Udio~\footnote{https://www.udio.com/} features a remix function that adjusts tones, allowing users to transform a song into a different musical style while preserving the original lyrics and main melody. However, these transformations are applied to the entire song, changing its overall style rather than making detailed adjustments to specific attributes or sections of the music. To address this limitation, MusicMagus~\cite{zhang2024MusicMagus} introduced a zero-sample text-to-music editing approach, enabling users to refine specific musical attributes using text-based prompts. 
However, their approach requires a certain level of musical expertise to achieve optimal results. Users need to know which attributes should be revised and how to frame them in prompts.

In this work, we aim to design a system that supports users in creating music using multimodal inputs. Through visual representations, users can understand how different elements are merged into the song, enabling them to edit music locally while preserving the overall melody and structure.

%% file: body/3-Research.tex
\section{Design Consideration}
To gain a deeper understanding of how professional musicians create and edit music, we conducted interviews with three target users: a remix creator (P1) with one year of experience, a rapper (P2) with four years of experience, and a self-published media blogger (P3) with three years of experience. The goal of these interviews was to gather diverse insights, expectations, and pain points related to AI music tools.

\begin{figure}[t]
    \centering
    \includegraphics[width=\linewidth]{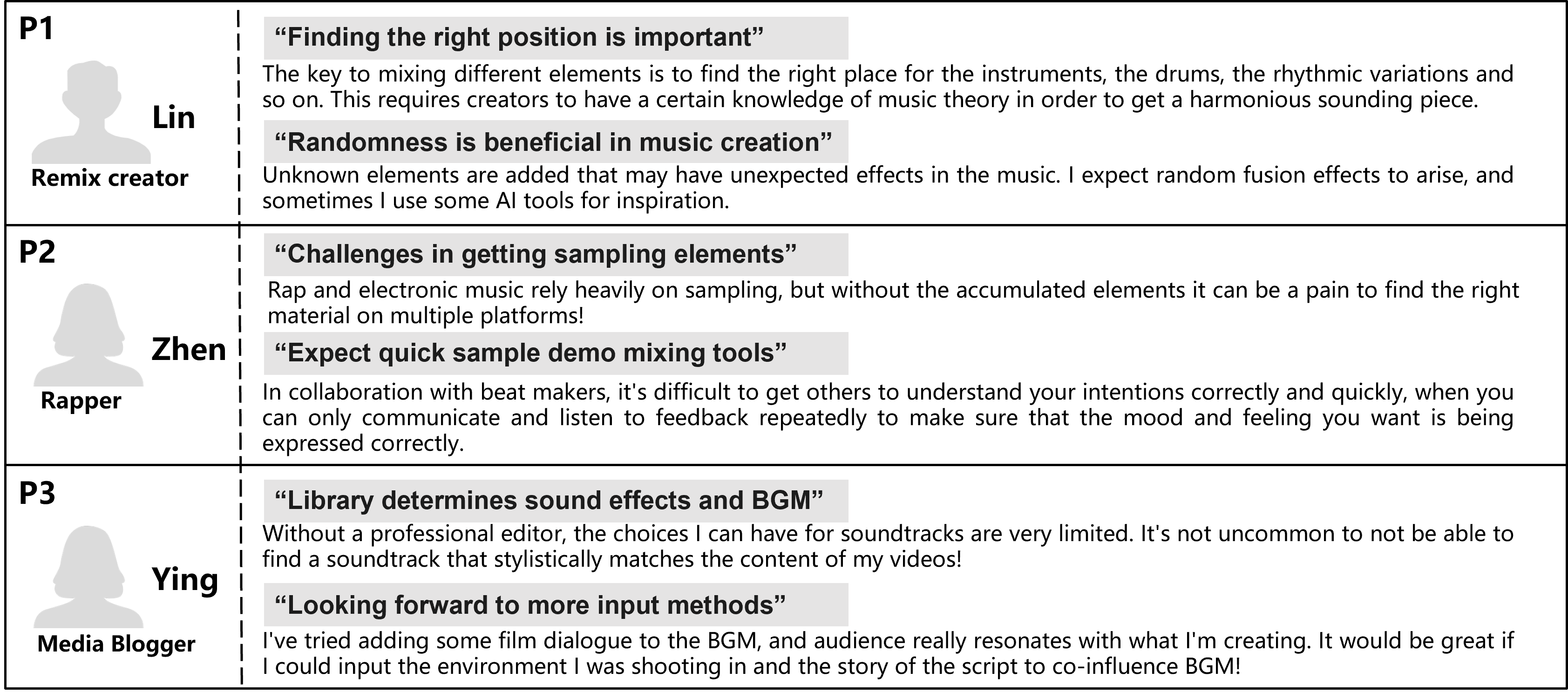}
    \caption{Suggestions and feedback from the target users.}
    \label{fig:requirements}
    \Description{The figure shows suggestions and feedback from the target users. On the left side, there is identifying information for three users who are, in order, a remix creator, a rapper, and a self-published media blogger. Their comments include, The two most central comments from each respondent have been selected on the right-hand side of the graph, in the order of 'Finding the right position is important', 'Randomness is beneficial in music creation', 'Challenges in getting sampling elements', 'Expect quick sample demo mixing tools', 'Library determines sound effects and BGM', 'Looking forward to more input methods'.}
\end{figure}

The interviews were semi-structured and lasted three hours, and were designed to allow participants to freely share their experiences and perspectives. The interviews began with open-ended questions about their music creation and editing processes, and then moved on to specific questions about their experiences with AI tools.
We gathered diverse insights into their opinions and needs (as summarized Figure~\ref{fig:requirements}). Our target users shared their approaches to creating and editing music, highlighted challenges in sourcing suitable sound materials, and emphasized the need for more diverse and efficient methods to assist in music generation.
Based on the interview results, we identified the following design requirements:
    \begin{itemize}[nosep,leftmargin=10pt]
        \item \textbf{Multimodal Input (P1, P2, P3).} The system should support diverse input modalities, including text, images, and music clips, to provide personalized pathways for conveying users' emotions and intentions. While each of these input types has been individually explored in music generation, the challenge lies in effectively combining them and integrating their information into the composition process. This design requirement stems from the belief that visual and auditory senses are inherently connected, as people create both images and music to express emotions. A well-integrated system should allow these modalities to merge seamlessly, enhancing the creative process. This perspective is also supported by prior studies on synesthesia \cite{Ward2008Synaesthesia, FILIMON2023Syncretism}, which suggest that the human brain naturally links multiple senses and emotions in response to a single stimulus. By leveraging multimodal input, the system can better align with human cognitive processes, offering a more intuitive and expressive approach to music creation compared to relying solely on single-modal inputs.
        \item \textbf{Visual Representations and Guidance (P2, P3).} The tool should offer visual representations showing how these inputs influence the generated music, along with guidance on where users can make changes or add elements. This feature is essential for supporting users with varying levels of expertise and accommodating different musical styles, helping them navigate and progress in their creative process.
        \item \textbf{Real-time Feedback (P1, P2).} The tool should enable real-time adjustments and previews, allowing users to modify parameters such as tempo, volume, and sound during music generation. Instant feedback on adjustments ensures that the music aligns with the user's creative vision, making the process more intuitive and efficient.
    \end{itemize}

%% file: body/4-SystemDesign.tex
\section{\mymethod}

We created \mymethod, a music creation and editing tool, as illustrated in Figure~\ref{fig:interface}. The interface is structured into two main sections. On the left side, users can search through a music library to select a foundational track and provide custom inputs, such as text, images, or audio files, to convey their creative intentions. The system then processes these elements to synthesize a personalized piece of music. The generated music is visualized and presented in the playback area on the right, where users can see how information extracted from diverse inputs is integrated into the composition.
A moving dot progresses along with the music playback, creating a bridge between what users perceive visually and what they hear aurally. 
This dual-section design allows users with little to no professional music knowledge to engage in the creative process, offering an accessible platform for musical expression. 

\begin{figure*}[t]
    \centering
\includegraphics[width=0.9\textwidth]{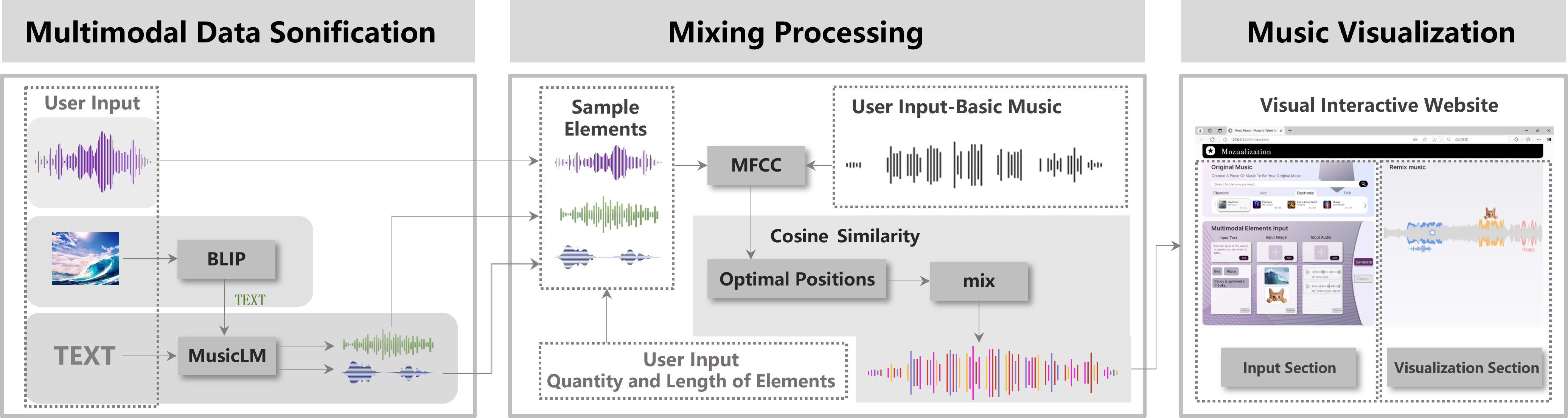}
    \caption{The workflow consists of three main components: multimodal data sonification, mixing processing, and music visualization. }
    \label{fig:framework}
    \Description{The workflow consists of three main components: multimodal data sonification, mixing processing, and music visualization. The first part shows text, images and audio converted to audio elements via BLIP and MusicLM. The second part shows the audio waveforms being mixed by MFCC and cosine similarity to determine the optimal mixing position and perform the mixing. The third part shows our user interface where the user can see the audio effect after visualization.}
\end{figure*}

As depicted in Figure~\ref{fig:framework}, the workflow of \mymethod consists of three main components: multimodal data sonification, mixing processing, and music visualization. 

\textbf{Multimodal data sonification.}
This component focuses on how information is extracted from user-provided text and images and translated into audio clips. The system processes user inputs by converting text and image data into audio representations for subsequent mixing.
For text data, color-related words are extracted and stored for mapping to the music visualization. For image data, the system employs the BLIP model~\cite{li2022blipbootstrappinglanguageimagepretraining}, which analyzes the image to extract key features and generates corresponding textual descriptions in natural language. For example, if a user uploads an image of a cat, the BLIP model may generate a description such as ``a yellow and white striped cat,'' from which the colors ``yellow'' and ``white'' would be extracted.

Thus, from both images and user-provided text, two key pieces of information are extracted: descriptive text and colors present in the image or mentioned in the text. The descriptive text is then transformed into audio clips using the MusicLM model~\cite{Agostinelli:2023:musiclm}, while the extracted colors are stored and mapped to the music visualization. This integration creates a seamless connection between visual and auditory elements, enhancing the coherence and expressiveness of the generated music.

\textbf{Mixing processing.}
This component focuses on determining where the generated audio clips should be integrated into the original music. To achieve this, the system analyzes each audio clip using the audio loading and beat tracking functions from the Librosa library~\cite{brian_mcfee-proc-scipy-2015}, enabling the extraction of fundamental parameters such as audio duration, pitch, and tempo.
Next, the system employs the Mel-Frequency Cepstral Coefficients (MFCC) method~\cite{davis1980comparison} for feature extraction, capturing essential acoustic characteristics of both the generated audio clips and the original music. To assess their compatibility, the system calculates the cosine similarity between the MFCC feature parameters of the audio segments. Cosine similarity measures the alignment between two feature vectors, with values ranging from $-1$ to $1$, where values closer to $1$ indicate a higher degree of similarity.
Using these similarity results, the system identifies optimal positions within the original music for integrating the generated audio clips. It then performs mixing operations at these points, ensuring a seamless and coherent audio composition that maintains the integrity of the original piece.

\begin{figure}[t]
    \centering        
    \includegraphics[width=\linewidth]{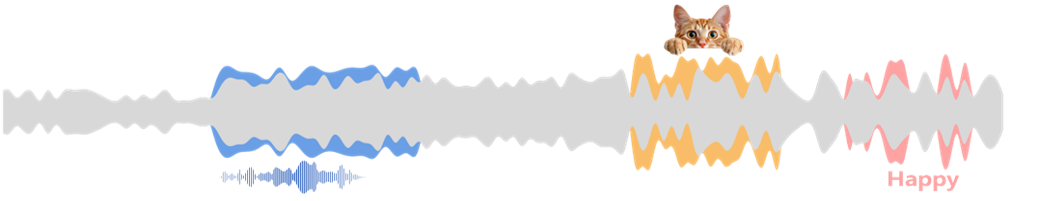}
    \caption{A stacked streamgraph visualization is used to represent the generated composition, with each layer corresponding to an input element. Preview images and text are embedded within the layers to enhance clarity and provide context.}
    \label{fig:visualisation}
    \Description{A stacked streamgraph visualization is used to represent the generated composition, with each layer corresponding to an input element. Preview images and text are embedded within the layers to enhance clarity and provide context. The figure shows the effect of each of the three input elements (from left to right: audio, image, and text depiction) after visualization, where they produce a stacked effect of audio at their respective mixing positions.}
\end{figure}

\textbf{Music visualization.}
This component shows where and how the extracted information is embedded into the original music. We developed a stacked streamgraph visualization to represent the generated composition, as shown in Figure~\ref{fig:visualisation}. Each input, such as keywords, text, or images, is represented as a layer in the streamgraph. The extracted color information is mapped to the layer's color, while the layer’s position indicates where the information is merged into the music. The thickness of the layer reflects the intensity of the added audio clip.
Additionally, input elements such as text or images are displayed near their corresponding layers for clarity. A moving dot progresses along the streamgraph in sync with the music playback, helping users connect what they see with what they hear.
This visualization enables users to understand how diverse inputs influence the original music, empowering them to make further edits. For example, users can upload new text or images and observe how both the music and visualization change. 

%% file: body/5-UserStudy.tex
\section{User Evaluation and Results}
We conducted a user evaluation to assess the usability of \mymethod and, more importantly, to gather feedback through interviews to identify areas for improvement in the next iteration of the design.

\subsection{Evaluation}
We invited nine music enthusiasts to use our prototype to create their own music and share their insights on the interface design and interaction techniques provided by the tool. These participants came from different majors at the university and all reported experience in searching for and listening to songs on various music applications. Among them, three had a background in music, with at least three years of experience in composition or other music-related creative work. The remaining six participants were regular music listeners with over five years of listening experience. They were familiar with various music genres and proficient in using different music apps, providing valuable perspectives from a listener's standpoint. The study received IRB approval for the protocol (XJTLU University Research Ethics Review Panel, \textnumero~20250112002310).

\begin{figure*}[t]
    \centering
    \includegraphics[width=0.95\textwidth]{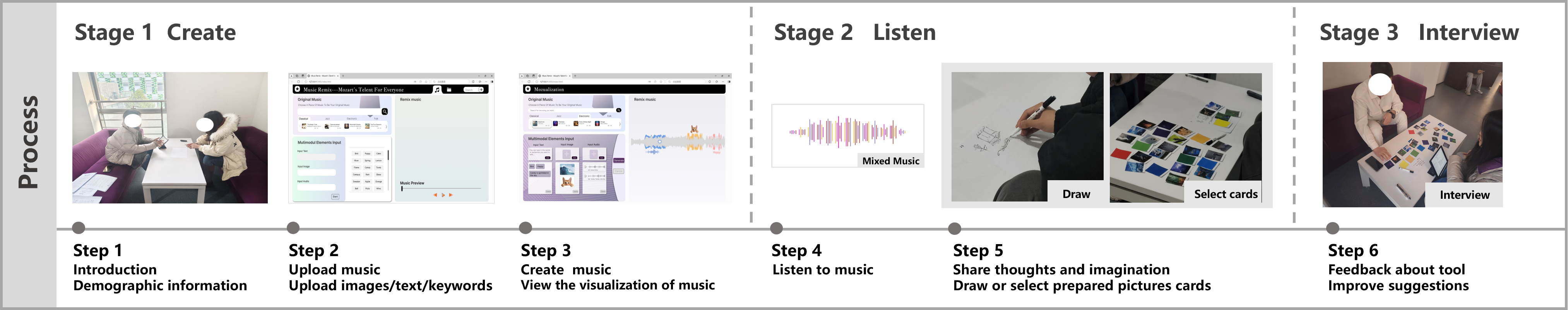}
    \caption{The evaluation consisted of three stages: create, listen, and interview.}
    \label{fig:evaluation}
    \Description{This figure shows an evaluation consisting of three stages: create, listen, and interview. On the left is the first stage, 'Create', which consists of steps 1, 2, and 3 and shows two people working at a table, one of them (the user) using a tablet and the other (the tester) holding a piece of paper, while the user learns about the functionality of the tool and then operates it with the help of the tester. In the middle is the second stage, 'Listen', which consists of steps 4 and 5, with step 4 showing a mix of music, and step 5 showing the user working on a drawing and card selection task. On the right is the third stage, 'Interview', which shows a person (the user) conducting an interview, with some cards and notes on the table, and he is pointing to the cards and describing them.}
\end{figure*}

As shown in Figure~\ref{fig:evaluation}, the evaluation process consisted of three stages:

The first stage aimed to evaluate the usability and user experience of our music generation tool. To begin, we introduced the goal of the evaluation and provided an overview of the tool. Participants were asked to share demographic information, such as their interests and experience with music applications. They were then instructed to choose a piece of music and upload it into the system. Additionally, we asked participants to upload images or input text/keywords to describe their mood or the elements they wished to incorporate into the music. After the music was generated, participants were invited to listen to it and view the visual representations, which demonstrated how their inputs were integrated into the result. They were encouraged to use the system to edit the music as needed. 

The second stage focused on users' experience in listening to the generated compositions, with a particular interest in whether they could recognize key messages derived from input images or perceive the joint compositions.
To assess this, we played a pre-generated piece of music and asked participants to listen carefully, observing their reactions and interpretations. While listening, participants were encouraged to share their thoughts and imagination regarding what they heard, as well as what they liked or disliked about the music. Furthermore, participants were invited to use colorful pencils to draw or select prepared pictures to visually represent how the music made them feel. All discussions, drawings, and selected pictures were recorded for further analysis.

In the third stage, participants were invited to use our tool further and create additional music. Following this, we conducted interviews to gather feedback about their experiences with the tool and their suggestions for improvement. The entire evaluation process lasted approximately one hour.

%% file: body/6-Result.tex
\subsection{User Experience in Music Creation}
Participants actively engaged with \mymethod and expressed a high level of satisfaction with its interface and functionality. The system was described as user-friendly and intuitive. The combination of text, images, and audio input facilitated flexible idea expression, allowing participants to explore and realize their creativity. Among the nine participants, eight had no prior knowledge in music composition and editing, while one had only engaged in basic music editing. However, following their interaction with \mymethod, all participants developed new insights into the process of music creation. They iteratively refined their compositions, experimenting with various input combinations to better align with their personal preferences and creative goals. In addition, all participants indicated a strong willingness to use \mymethod again. They noted that the system's interactive stimulated their curiosity and fostered a sense of engagement, increasing their motivation to use the tool in the future.

\subsection{Perception of Generated Music and Visualization}
Participants provided positive feedback on the generated music, noting that \mymethod effectively facilitated the fusion of musical components, resulting in a cohesive and engaging auditory experience. All participants found it exciting to identify the elements they had provided and were intrigued by how these inputs were transformed into musical components and seamlessly merged with others. They appreciated the system’s ability to integrate diverse elements into harmonious compositions, even when the individual components differed significantly in style. For instance, a peaceful melody and Peking Opera were successfully merged with complementary rhythm and placement. One participant expressed initial skepticism about incorporating bell chimes, fearing they might create a harsh or discordant effect. However, after listening to the generated composition, he noted that the bell chimes enriched the auditory experience: ``The appearance of the bell with the music feels like someone is praying,'' perfectly aligning with the narrative conveyed by the music. Additionally, three participants highlighted that combining real natural sounds with the music elicited stronger emotional responses. They emphasized that this approach enhanced the music’s ability to evoke imagery, creating an immersive and emotionally resonant experience for listeners.

Participants also provided positive feedback on the music visualization, emphasizing its utility in enhancing their understanding of the generated compositions. By observing waveform changes and input elements, they were able to better comprehend how the diverse elements were integrated into the final output. Participants noted that this feature provided an intuitive way to connect auditory and visual elements. The visualization not only enriched the music's sense of imagery but also effectively conveyed the musical qualities of the text and images, further deepening user engagement with the creative process.

\subsection{Improvement Suggestions}
Some participants provided constructive feedback for enhancing \mymethod to better support the generation of high-quality musical compositions. One participant recommended implementing measures to avoid frequency masking during audio mixing, which occurs when two sounds occupy the same frequency range, leading to their mutual concealment and a reduction in the clarity of the final mix.
Moreover, two participants noted that when blending a melodic audio clip with another piece of music, a more nuanced mixing processing approach might be required to coordinate the tuning discrepancies between them. Such adjustments would help ensure a smoother integration of the audio components, further improving the overall cohesiveness of the generated music.

%% file: body/7-Discussion.tex
\section{Discussion and Future directions}
\mymethod represents an initial design for generating customized music using diverse inputs. Our concept is that text, images, and audio clips serve as powerful mediums for expressing emotions, each carrying a unique ``color'' and ``sound'' when abstracted from their original forms. By encouraging innovative ways to merge these inputs, \mymethod enables the creation of outputs that effectively convey users' emotions. whether in a musical or visual form. 
Visualization, in particular, plays a crucial role in showing how and where diverse inputs are integrated into the music. This not only helps users understand the generated results but also provides a platform for further refinement. Users can edit a wide range of variables, including the color, shape, position, and texture of the visualization, as well as the tone, rhythm, timbre, and volume of the music. These customized features open up opportunities for future exploration and innovation in creative expression. 
Such versatility also makes \mymethod suitable for a wide range of applications. For instance, in art galleries, it can generate music to complement exhibit pictures, enhancing the viewer's experience. In opera, it can create background visuals that immerse the audience in the narrative, making performances more engaging and emotionally impactful.

While \mymethod enhances music creation, it has limitations that require further exploration. One key challenge is the conversion of images into textual descriptions, which may lead to a loss of detail and reduce their expressive potential in music composition. Additionally, the current editing capabilities lack precision, making fine-grained adjustments and complex creative expression difficult.
This limitation was also highlighted in the user study, where participants struggled to accurately place music clips at their desired positions. The main reason for this difficulty is that our system relies on AI to determine the optimal placement of joint music clips. While the generated results sound cohesive, and the visualization effectively illustrates when and where elements are embedded, users expressed a desire for more control---specifically, the ability to select an approximate time location for embedded music content.
To address these challenges and support more intuitive music creation, we propose three future research directions:

\textbf{Enhancing Visualization Editing Capabilities.} It is necessary to enhance the current visualization editing functionality to support fine-grained control over added elements. For example, adjusting the height of elements in the visualization could modify pitch, while altering their horizontal position could fine-tune their timing within the composition. Additionally, users might wish to make localized adjustments, updating the melody of particular sections. These advanced editing capabilities would enable users to explore and refine their creative ideas with greater precision, striking a balance between algorithm-driven generation and user-driven customization. Such functionality would be particularly valuable for users with prior musical knowledge, allowing them to leverage their expertise to achieve more sophisticated and nuanced \mbox{compositions}. 

\textbf{Extending to Immersive Environments.}
Expanding the system to immersive environments, such as VR or AR, could create a more engaging and sensory-rich platform for music creation. In these environments, music elements could be designed as virtual objects that users can place, move, and manipulate within 3D space, allowing users to compose music in a dynamic way. Additionally, immersive environments could enable users to experience their compositions from a variety of perspectives, moving through the virtual space to hear how different arrangements and spatial placements of musical elements affect the overall sound, fostering exploration and discovery in ways that traditional 2D interfaces cannot replicate.  

\textbf{Expanding Input Modalities.} Incorporating additional input modalities could greatly enhance user interaction within the music generation process, particularly in immersive environments. Unlike in 2D interfaces, where editing is limited, users themselves could become active control points in the generated music. For example, users' positions, movements, and gestures could dynamically influence the playback, such as adjusting volume or tempo as they move through the space or interact with virtual objects. This would enable a more embodied and intuitive way of editing and personalizing the music, deepening the connection between the user and their creation.

%% file: body/8-Conclusion.tex
\section{Conclusion}
In this work, we developed \mymethod, a multimodal music generation tool designed to inspire music and art enthusiasts to explore innovative approaches to their creations. We believe this tool has significant potential for refinement, with its resulting innovations applicable to a variety of scenarios. By delving into future improvements and research directions, we can further advance multimodal generation technology, unlocking new opportunities and fostering greater innovation in the fields of music and art creation.